# A facile vector substrate platform via BaTiO$_3$ membrane transfer enables high quality solution processed epitaxial PZT on silicon

Asraful Haque*, Antony Jeyaseelan, Shubham Kumar Parate, Srinivasan Raghavan, Pavan Nukala*.
Centre for Nanoscience and Engineering, Indian Institute of Science, Bengaluru, India, 560012.

**Abstract**
The direct integration of high-performance ferroelectric oxides with silicon remains challenging due to lattice mismatch, thermal incompatibility, and the need for high-temperature epitaxial growth. Here, a hybrid integration approach is demonstrated in which crystalline BaTiO$_3$ (BTO) membranes are first transferred onto Pt-coated Si substrates and subsequently used as vector substrates (VS) for the growth of epitaxial (001) Pb(Zr$_{0.52}$Ti$_{0.48}$)O$_3$ (PZT) thin films via chemical solution deposition (CSD). A KI + HCl-based etchant enables rapid and complete dissolution of the SrVO$_3$ sacrificial layer in about 30 minutes, reducing the release time from days to minutes compared with conventional water-based approaches to dissolve AVO$_3$/AMoO$_3$ (A = Ca, Sr, Ba). The BTO VS imposes dominant (00l) out-of-plane orientation and in-plane cube-on-cube epitaxy in the overlying PZT. Devices exhibit P$_r$ ~ 10-12 µC/cm$^2$ and E$_c$ ~ 100 kV/cm, with stable switching to 10$^8$ cycles on the VS. From piezoelectric butterfly loops we extract $d_{33}^{effective}$ ~ 70 pm/V for PZT on VS), ~ 54 pm/V for PZT grown on conventional Pt–Si substrates. This approach demonstrates a scalable and cost-effective route for integrating functional ferroelectric materials onto silicon and offers a promising platform for future CMOS-compatible oxide electronics.

1. Introduction

The integration of dissimilar materials through heteroepitaxy, where a crystalline film grows with a well-defined out-of-plane orientation on a chemically and structurally distinct substrate, is central to the development of multifunctional electronic and photonic devices [1]. However, achieving high-quality heteroepitaxial growth remains challenging due to intrinsic mismatches in lattice constants, thermal expansion coefficients, and interfacial chemistries. These mismatches often lead to high defect densities, interfacial reactions, and compromised film quality; challenges that are particularly pronounced when integrating complex oxides with silicon (Si). While traditional approaches to remove the native Si oxide, such as wet chemical treatments [2] and high-temperature annealing [3], have been widely used, more advanced methods rely on ultrahigh vacuum (UHV) based strontium (Sr) or strontium oxide (SrO) deoxidation. These techniques enable not only effective oxide removal but also atomic-scale interface engineering, allowing the formation of a Sr-reconstructed Si surface. Historically implemented via molecular beam epitaxy (MBE) [4] and atomic layer deposition (ALD) [5], such strategies have recently been extended to pulsed laser deposition (PLD), using Sr [6] or SrO [7] sources to enable in situ deoxidation.

Among functional oxides, ferroelectric and piezoelectric materials such as Pb(Zr$_x$Ti$_{1-x}$)O$_3$ (PZT) are of particular interest due to their large piezoelectric coefficients [8], high electromechanical coupling [9], and switchable polarization especially near the morphotropic phase boundary [9]. These properties make them attractive for microelectromechanical systems (MEMS) [10], sensors, actuators, and energy harvesters [11]. To fully leverage their functionalities in integrated devices, it is essential to maintain their single-crystalline nature when interfaced with Si. However, direct epitaxial growth of PZT on Si remains difficult due to its poor interfacial chemical compatibility and the high temperatures typically required for oxide synthesis [12]. Consequently, integration strategies often rely on complex buffer architectures

(e.g., PZT/SRO/CeO$_2$/YSZ//Si [13], PZT/SrRuO$_3$/Pt/ZrO$_2$//Si [14], PZT/LSMO/STO//Si [15]) or alternative platforms such as graphene [16], which can support highly oriented oxide growth, but present challenges related to thermal stability and interface protection during deposition [17]. Chemical solution deposited (CSD) PZT on Pt/Ti/Si typically forms an oriented or polycrystalline rather than epitaxial, with the interface chemistry biasing texture (TiO$_2$-rich surfaces favour (111), whereas Pt-rich surfaces promote (001)/(100)) [18]. The piezoelectric response is strongly anisotropic, so among common orientations the (001) rhombohedral variant provides the largest out of plane coefficient [19]. Accordingly, achieving (001)-textured PZT directly on Si is a key lever for high-performance piezo-MEMS.

An emerging and versatile approach is layer transfer, in which single-crystalline oxide films are grown on lattice-matched oxide substrates using a sacrificial buffer layer, and then released and transferred onto Si. Water-soluble sacrificial layers such as Sr$_3$Al$_2$O$_6$ (SAO) [20], SrVO$_3$ (SVO) [21], SrO [22], and BaO [23] have been widely studied for this purpose. While SAO and similar compounds offer fast dissolution, their poor atmospheric stability complicates handling and storage [24]. SVO, on the other hand, is considerably more stable under ambient atmospheric conditions and compatible with high-quality oxide growth yet suffers from slow dissolution kinetics, often requiring several days to fully release freestanding films [21,25].

To address this bottleneck, our earlier work focused on accelerating SVO dissolution through strategies such as monolayer graphene implantation and top film patterning [21]. Building on that, in this study we show that a KI + HCl redox etchant reduces the SVO dissolution time from days to about 30 minutes while preserving BaTiO$_3$ (BTO) crystallinity. The released BTO membranes are transferred onto Pt–Si and used as vector substrates (VS) [26] for the chemical-solution deposition of epitaxially aligned (001) PZT thin films. This VS-enables CSD route combines room-temperature transfer with moderate temperature (650 °C) PZT crystallization on Si, avoiding high temperature epitaxy. The templated PZT exhibits dominant (00l) out-of-plane orientation and in-plane cube-on-cube epitaxy (fourfold φ-scan of PZT {022}). Devices show robust ferroelectric switching with stable remanent polarization to $10^8$ cycles and a stronger longitudinal electromechanical response than PZT grown directly on Pt–Si. The combination of rapid SVO lift-off, VS templating, and CSD processing provides a practical pathway for CMOS-compatible oxide ferroelectrics and piezoelectrics on silicon.

2. Results and Discussion

2.1 Atmospheric Stability of SVO vs SAO

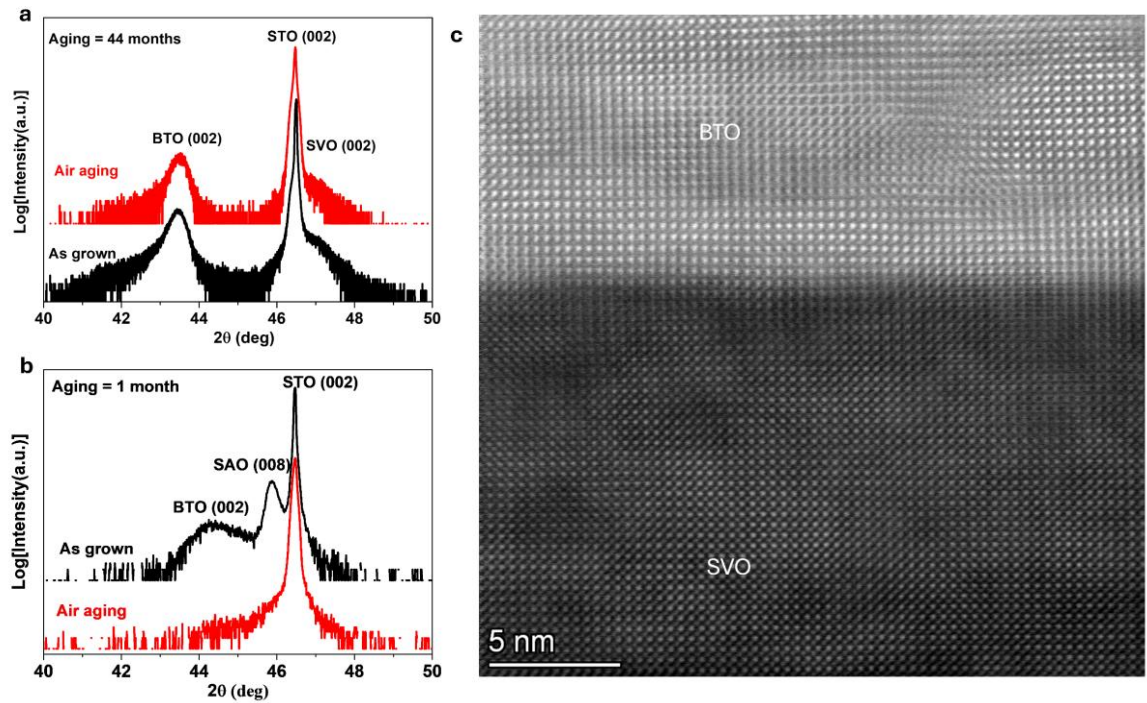

**Figure 1: Structural characterization demonstrating ambient atmosphere stability of SVO over SAO sacrificial layers grown on STO.**
(a) X-ray diffraction of the BTO/SVO//STO heterostructure measured immediately after growth and after 44 months of air aging. (b) X-ray diffraction of the BTO/SAO//STO heterostructure measured after deposition and following one month of atmospheric aging. (c) Cross-sectional atomic-resolution HAADF-STEM image of the BTO/SVO//STO heterostructure after 44 months of atmospheric aging.

To evaluate the atmospheric stability of SVO, we compared it directly with that of SAO under identical processing conditions. Epitaxial SVO thin films were grown on single-crystalline (001) $SrTiO_3$ (STO) substrates via pulsed laser deposition (PLD) using a polycrystalline SVO target at a laser fluence of 1.4 J/cm², a base pressure of $3 \times 10^{-6}$ mbar, and a substrate temperature of 750 °C. Without breaking vacuum, a BTO capping layer was subsequently deposited on top of the SVO using a polycrystalline BTO target at 1.2 J/cm², $5 \times 10^{-6}$ mbar, and 800 °C. For comparison, SAO films were also grown on STO under a laser fluence of 1.3 J/cm², at 800 °C and $5 \times 10^{-6}$ mbar, followed by BTO deposition under the same conditions used above. The BTO/SAO//STO heterostructure was subsequently annealed at 900 °C.

Figure 1a shows the symmetric θ–2θ high resolution X-ray diffraction (XRD) scan of the as-grown BTO/SVO//STO heterostructure in the 40°–50° range, exhibiting well-defined Bragg peaks from BTO (002), SVO (002), and the STO substrate (002), confirming the out of plane orientation of the grown film. Remarkably, even after 44 months of storage under ambient atmospheric conditions, these Bragg reflections remain clearly visible in the θ-2θ scan, indicating excellent long-term stability of both SVO and BTO. After atmospheric aging, a cross-sectional FIB lamella of the BTO/SVO//STO stack was prepared and imaged by high angular annular dark field scanning transmission electron microscopy (HAADF-STEM). The

HAADF-STEM image acquired along the [001] zone axis of the STO substrate is shown in Figure 1c, revealing a sharp and coherent interface between BTO and SVO. The crystalline quality of both layers remains intact even after 44 months of ambient exposure, highlighting the exceptional long-term structural stability of the SVO-based heterostructure.

In contrast, the BTO/SAO//STO heterostructure exhibited significantly poorer atmospheric stability. As shown in Figure 1c, the as-grown high resolution X-ray diffraction θ–2θ scan displays Bragg peaks corresponding to BTO (002), SAO (008), and STO (002), confirming successful deposition. However, after just one month of exposure to ambient conditions, the SAO peak completely disappears and the BTO (002) peak becomes significantly diminished, clearly indicating degradation of the SAO layer and loss of film integrity. These results underscore the superior ambient stability of SVO compared to SAO, establishing SVO as a more robust sacrificial layer for heterostructure transfer and long-term storage.

## 2.2 Accelerated dissolution of SVO sacrificial layer

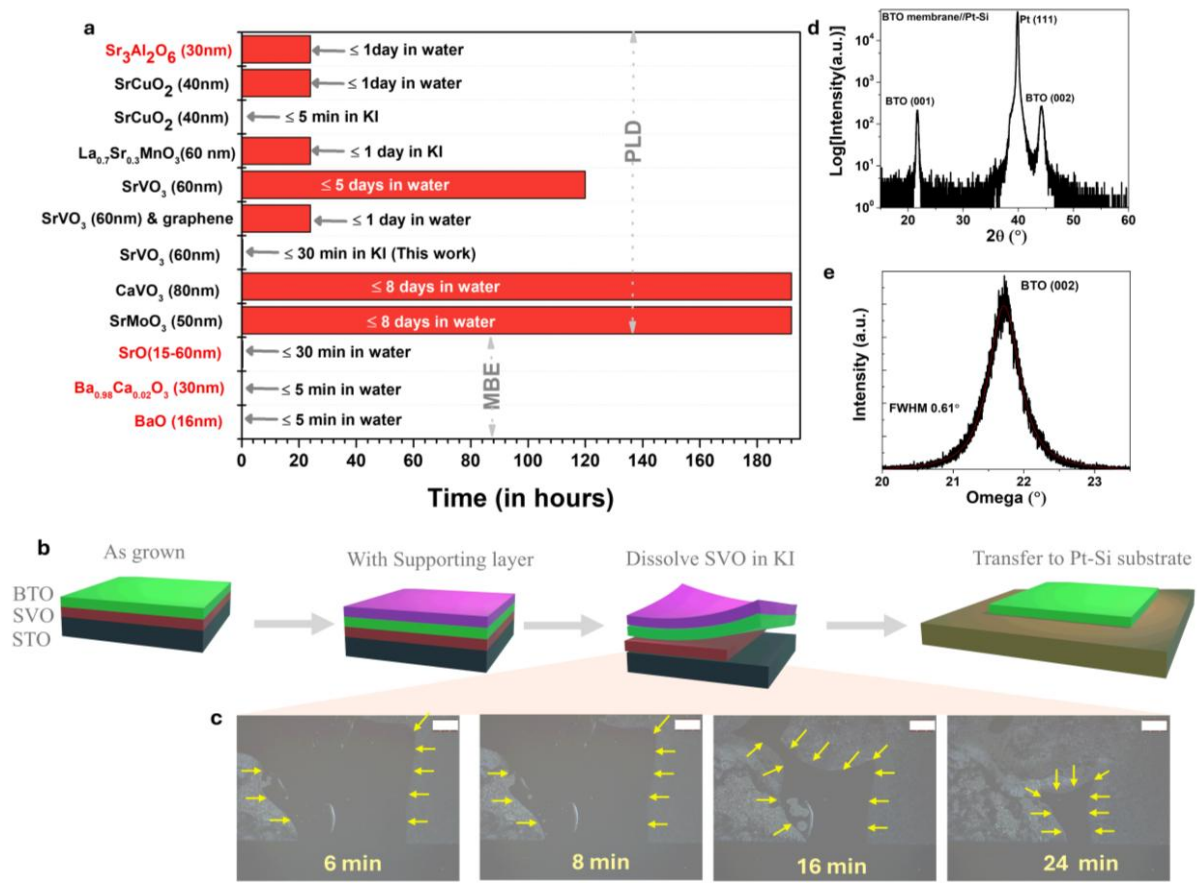

**Figure 2: Exfoliation, transfer, and characterization of BTO membranes.** (a) Comparison of the etching rate of SVO thin films in KI solution versus deionized water, along with benchmarking against other sacrificial layers reported in the literature [21,22,24,25,27,28,29,23]. (b) Schematic representation of the membrane transfer process using KI + HCl solution. (c) Optical image of the BTO/SVO//STO sample immersed in KI + HCl solution; the yellow arrow indicates the progressing SVO dissolution front. Scale bar: 250 μm (top right of each optical

image). (d) High resolution X-ray diffraction (θ–2θ) scan of the transferred BTO membrane on Si. (e) Rocking curve around the BTO (002) reflection for the transferred membrane on Pt-coated Si.

While atmospheric stability is essential for storage and handling, efficient and controlled dissolution of the sacrificial layer is equally critical for the successful release of freestanding membranes. Figure 2a presents the dissolution times of various sacrificial layers in both deionized water and potassium iodide (KI) solution. Sacrificial materials with poor atmospheric stability; including SAO, SrO, BaO, and $BaCO_3$ are highlighted in red on the vertical axis. These compounds, particularly the alkaline earth metal oxides, readily undergo hydrolysis in the presence of moisture, following the general reaction:

$$MO + H_2O \rightarrow M(OH)_2$$

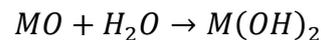

Where *M* represents a Group II metal (e.g., Sr, Ba).

Among all the water-soluble sacrificial layers, SAO has gained popularity due to its fast dissolution rate in water. Although BaO and SrO dissolve even more rapidly, their incorporation in pulsed laser deposition (PLD) processes is complicated by the hygroscopic nature of the targets, which necessitates stringent control over ambient humidity and careful storage conditions. Furthermore, while SAO enables quick membrane release, it suffers from poor atmospheric stability, as discussed earlier.

In contrast, other water-soluble perovskite-type sacrificial layers such as SVO [25], $CaVO_3$ (CVO) [28], and $SrMoO_3$ (SMO) [29] exhibit much greater atmospheric stability, making them more suitable for long-term storage and handling. However, their dissolution kinetics in water are significantly slower, often requiring 5-8 days for complete lift-off, as shown in Figure 2a. To address this limitation, our previous work demonstrated that the dissolution rate of SVO in BTO/SVO//STO heterostructures can be enhanced by implementing a remote epitaxy based approach by utilizing graphene at the interface of SVO and STO [21]. This behaviour is attributed to an enhanced surface diffusion mechanism in graphene-like layered materials, whereby nanodroplets of water surf through propagating ripples [30]. In general, rapid lift-off of chemically stable oxides relies on acidic and/or redox-assisted etching. Sacrificial perovskites such as $La_{0.7}Sr_{0.3}MnO_3$ (LSMO) and $SrCoO_3$ (SCO) have been released using strong acids and iodide/iodine chemistry [31,27]. (Fig. 2a). By analogy, we employ a KI + HCl redox assisted etch for SVO: the acidic medium and the $I^-/I_2$ couple together accelerate dissolution. Because vanadium speciation depends on pH, chloride activity, and redox potential, we do not assert a single global reaction (refer to mechanistic notes in the SI). This KI-assisted route reduces the SVO etch time from days to about 30 minutes, enabling rapid release of freestanding BTO membranes.

## 2.3 Transfer process and structural characterization of transferred BTO membrane

Building on the enhanced dissolution strategy for SVO, we employed a KI+HCl solution to rapidly release ~150 nm BTO membrane from the BTO/SVO//STO heterostructure. As shown in Figure 2b, the sample was first coated with a polymethyl methacrylate (PMMA) polymer support layer to maintain the structural integrity of the freestanding film during transfer. Upon immersion in the KI+HCl solution, the SVO sacrificial layer dissolved efficiently, allowing the BTO membrane to be cleanly separated and transferred onto a Pt-coated Si substrate. The progress of the etching front was tracked in real time under an optical microscope (Figure 2c), revealing uniform and complete dissolution within 30 minutes; significantly faster than the >5 days duration typically required in water (Figure 2a).

Post-transfer structural analysis by X-ray diffraction confirmed that the crystalline quality of the BTO film was preserved. As shown in the symmetric θ–2θ scan (Figure 2d), the (001) and (002) peaks of SVO were completely absent, indicating full removal of the sacrificial layer, while the BTO (001) and (002) peaks remained sharp and well-defined. This confirms that the out-of-plane crystallographic orientation of the transferred BTO membrane is retained from the as-grown heterostructure. Additionally, the full-width at half maximum (FWHM) of the rocking curve for the BTO (002) peak in the transferred film was measured to be 0.61° (Figure 2e), closely matching the value of 0.43° for the as-grown film (Figure S1), further indicating minimal degradation in crystalline quality during transfer.

**2.4 Crystallization of PZT on a BTO vector substrate via chemical solution deposition (CSD)**

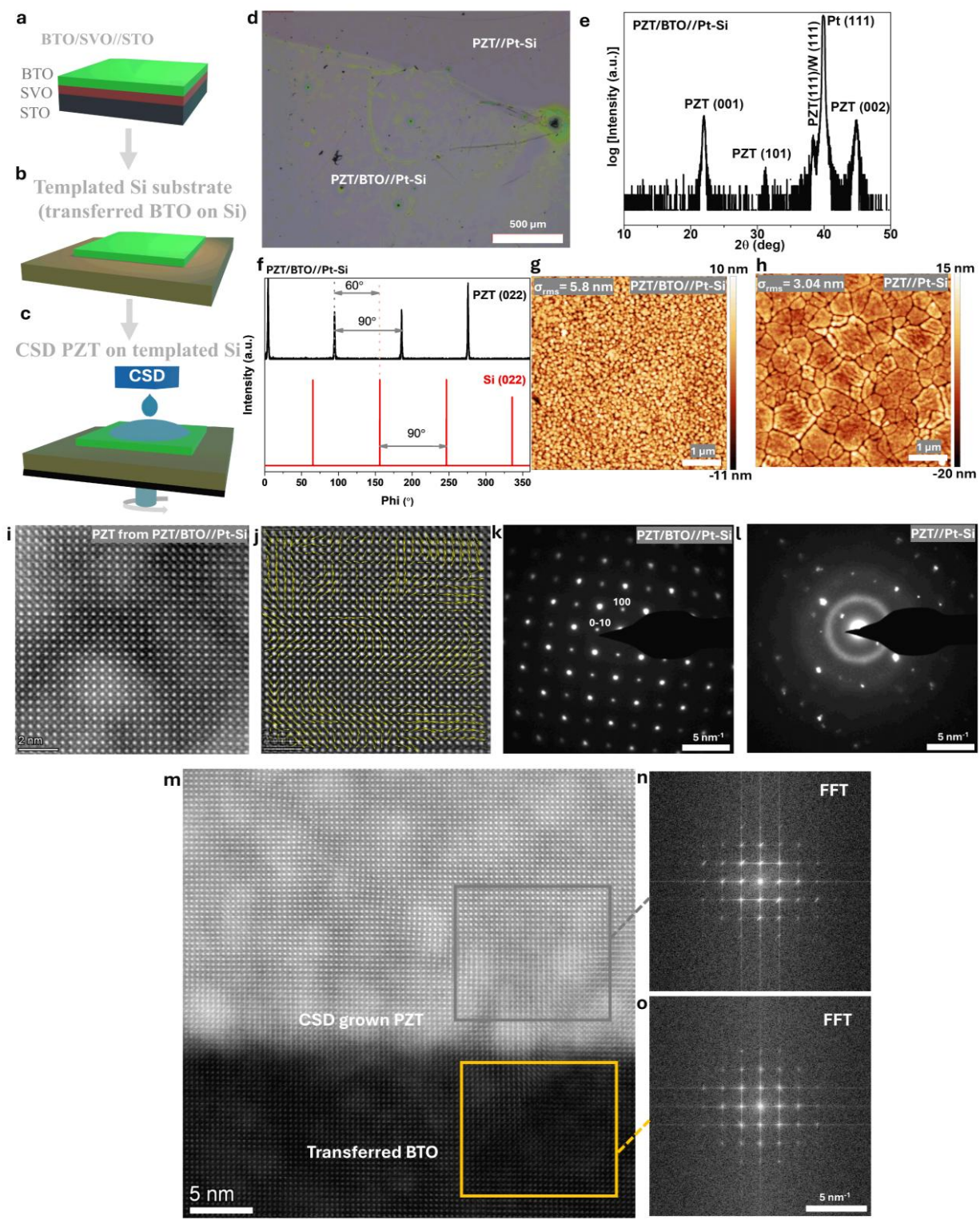

**Figure 3. Structural and microstructural characterization of PZT films grown via chemical solution deposition (CSD) on VS and directly on Pt–Si.**

(a–c) Schematic illustration of the heterogenous integration of PZT on VS: (a) Epitaxial BTO/SVO heterostructures are grown on STO substrates, (b) the BTO membrane is released and transferred to a Pt–Si substrate to obtain VS, and (c) PZT films are deposited via CSD on the VS. (d) Optical microscope image of the sample showing the spatially defined regions of PZT grown on VS (PZT/BTO//Pt–Si) and PZT//Pt–Si. (e) High resolution X-ray diffraction (XRD) θ–2θ scan of PZT grown on the VS. (f) Phi scan of the PZT (022) reflection for the PZT grown on VS showing fourfold symmetry, confirming in-plane crystallographic alignment with the underlying BTO template and epitaxial registry with the Si (022) reflection. (g,h) Atomic force microscopy (AFM) topography images of PZT films grown on VS and on Pt-Si substrate. (i,j) Atomic-resolution HAADF-STEM image and corresponding polarization vector mapping of the PZT film on VS, revealing nanoscale polar ordering and spatial variation in dipole orientation. (k,l) Selected-area electron diffraction (SAD) patterns from PZT grown on VS and PZT grown on Pt–Si regions, respectively. (m) Cross-sectional HAADF-STEM image of the PZT grown on VS, showing the interface between the CSD-grown PZT layer and the transferred BTO membrane. (n,o) FFTs from the PZT and BTO regions marked in (m)

Next, we applied the KI + HCl redox-release to fabricate a BTO vector substrate (VS): a ~30 nm epitaxial BTO membrane transferred onto a Pt–Si substrate. This VS was then used as the crystallographic template for chemical-solution deposition of epitaxial (001) $Pb(Zr_{0.52}Ti_{0.48})O_3$ (PZT), as schematically shown in Figures 3a–c. The detailed sol–gel synthesis is reported elsewhere [32]. Briefly, the PZT sol was spin-coated at 4000 rpm for 30 s, followed by a two-step pyrolysis (first at 400 °C for 2 min, then rapid thermal annealing at 650 °C for 2 min) and a final anneal at 650 °C for 15 min to ensure crystallization

Figure 3d shows the optical microscope image of the transferred BTO layer on Pt-Si (schematic cross-section shown in Figure 3b). Figure 3e presents the symmetric high resolution X-ray diffraction scan of the PZT grown on VS, revealing a predominant (00l) orientation of PZT. Minor reflections corresponding to the (101) and (111) planes are also observed, which are attributed to PZT grown directly on Pt-Si outside the BTO-templated region.

To probe the in-plane orientation, medium resolution XRD φ-scans were performed on the (022) asymmetric Bragg reflection of PZT, as shown in Figure 3f. The appearance of four-fold symmetric (022) peaks at 90° intervals confirms the retention of in-plane crystallographic orientation of PZT. A slight angular offset between the PZT (022) and Si (022) reflections is observed, which is attributed to the tilt introduced during the BTO membrane transfer, resulting in a misalignment between the BTO [100] and Si [100] directions. Atomic force microscopy (AFM) shows that PZT on the VS exhibits smaller lateral surface features and a lower RMS roughness (~3.04 nm) than PZT grown directly on Pt–Si (~5.8 nm) (Fig. 3g, h). The smoother surface is consistent with epitaxial templating that promotes oriented nucleation and reduces coalescence roughening. This trend mirrors reports on epitaxial $BiFeO_3$ thin films, where epitaxy yields denser, smoother morphology than polycrystalline counterparts [33].

Figures 3i and j display the atomic-resolution HAADF-STEM image of the PZT layer grown on VS and the corresponding polar displacement vector map. The overlaid polarization vectors in Fig. 3j show polar displacement fields, with smoothly varying orientations that indicate the formation of nanoscale ferroelectric domains. The selected-area diffraction (SAD) pattern from the PZT layer grown on the VS (Figure 3k) exhibits a sharp, periodic array of diffraction spots along a distinct (100) zone axis. The absence of rings and the fourfold symmetry of the pattern indicate strong out-of-plane (00l) texture and in-plane cube-on-cube registry with the underlying single-crystalline BTO. This confirms epitaxial growth of (001) PZT on the BTO VS. In contrast, the SAD pattern of PZT directly grown on Pt–Si (Figure 3l) displays broad, continuous diffraction rings, indicative of a polycrystalline film with oriented grains. This difference reflects the lack of lattice-matched guidance during crystallization on Pt–Si. Figure 3m presents a cross-sectional STEM-HAADF image encompassing both the PZT and BTO layers. The contrast difference between the top and bottom layers reflects the higher average atomic number of the Pb-containing PZT relative to BTO. The PZT/BTO interface appears sharp and well-defined, with no observable amorphous interlayer or contrast gradient, indicating a structurally coherent interface. The underlying BTO membrane retains uniform contrast and crystallinity, confirming its structural integrity following both the transfer and subsequent PZT growth.

**2.5 Ferroelectric and Electromechanical properties of PZT grown on VS and Pt-Si devices.**

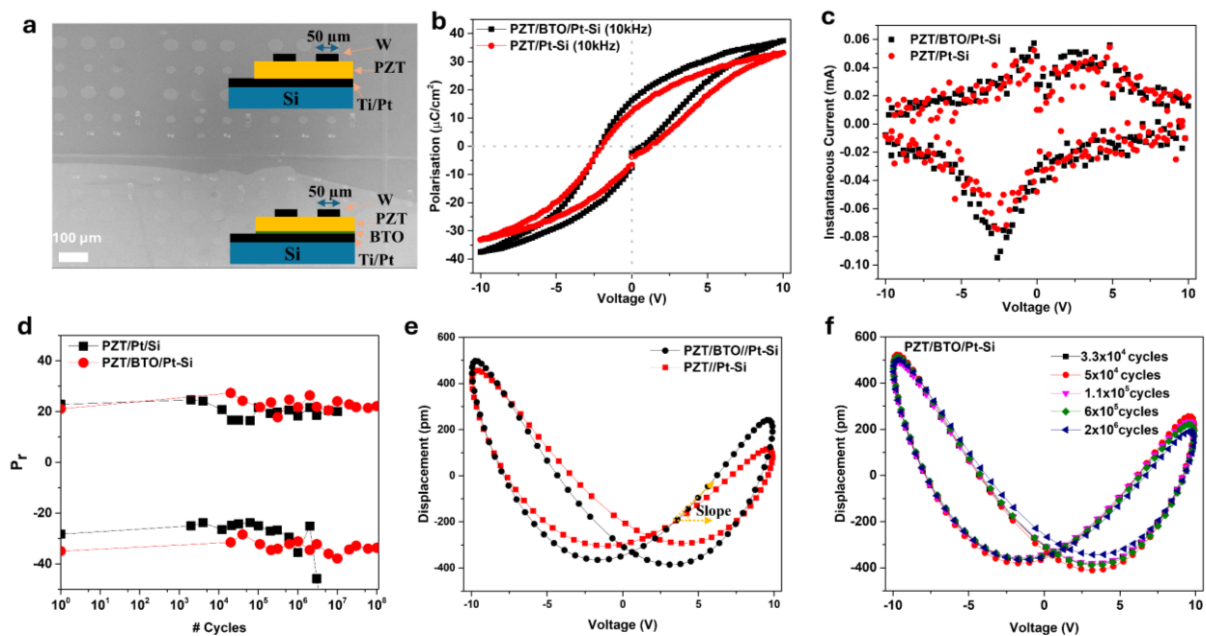

**Figure 4: Electrical and electromechanical characterization of PZT films grown via CSD on VS and Pt-Si** (a) SEM image of circular W electrodes fabricated on the PZT grown on VS and Pt-Si regions; inset shows the schematic device structure. (b) Polarization-voltage hysteresis loop measured at 10 kHz. (c) Instantaneous current (I) vs voltage (V) response corresponding to polarization switching. (d) Endurance behaviour of the remanent polarization ($P_r$), over $10^8$ switching cycles for PZT VS. (e) Displacement-voltage butterfly loop measured using laser Doppler vibrometer. (f) Displacement response under electrical cycling, showing stability over $10^6$ cycles for PZT grown on VS.

For electrical characterization, tungsten (W) circular electrodes were patterned on both PZT grown on VS and PZT grown on Pt-Si heterostructures using standard photolithography followed by metal deposition. Figure 4a shows an SEM image of the fabricated W electrodes, with a schematic cross-section provided in the inset. Ferroelectric measurements performed at 25 °C and 10 kHz reveal distinct polarization switching peaks in the instantaneous current vs voltage curves and well-defined hysteresis loops in the polarization vs voltage response (Figure 4b). The remanent polarization ($P_r$) and coercive field ($E_c$) are 10-12 µC/cm$^2$ and ~100 kV/cm, respectively (Fig. 4b,c). The $P_r$ values are comparable to previously reported (001) textured

PZT films on Pt-Si substrates. The higher Ec, relative to typical ~ 40-80 kV cm$^{-1}$ reported for ~ 250–300 nm PZT on Pt/Si, is consistent with the different mechanical/electrical boundary conditions of the VS architecture and series voltage partitioning across the non-ferroelectric thin BTO underlayer [24]; together with the 10 kHz measurement.

Device endurance was evaluated by applying rectangular voltage waveforms at 10 kHz and intermittently recording polarization using the PUND method. As shown in Figure 4d, the devices exhibited exceptional stability with no evidence of wake-up or fatigue over $10^8$ switching cycles for PZT grown on the VS. However, for PZT on Pt–Si the capacitor fatigues beyond ~$10^6$ cycles. This divergence despite using the same W top electrode implicates the bottom interface and film microstructure as the dominant factors. Polarization fatigue in PZT is widely understood to be interface-controlled, arising from the growth of a thin non-ferroelectric interfacial layer and charge-injection/oxygen-vacancy drift that pin domain walls [34,35,36] Direct evidence shows that eliminating the defective interfacial dielectric (e.g., by using oxide/oxide contacts) yields fatigue-free behaviour over $10^9$–$10^{12}$ cycles, whereas metal/PZT contacts on Pt/Si commonly degrade near $10^6$–$10^8$ cycles depending on interface quality [37–39]. These mechanisms are consistent with our results; the perovskite BTO template provides a chemically compatible bottom interface and promotes epitaxy, suppressing interfacial layer formation and injection, so $P_r$ remains stable to $10^8$ cycles, unlike the Pt–Si case.

The electromechanical response of the devices was further investigated by applying a large-amplitude sinusoidal AC voltage while monitoring out-of-plane displacement using a dual-beam laser Doppler vibrometer. Figure 4e displays the resulting butterfly-shaped voltage–1templated Si and PZT on Pt-Si. A phase lag is observed between the applied voltage and the resultant displacement response, leading to a non-zero displacement at zero voltage, as depicted in Figure 4e. The upturns of the butterfly loops coincide with the coercive voltage, measured from peak positions of the instantaneous current versus voltage curves, typical of the electrostrain response of a ferroelectric material. From these data, an effective piezoelectric coefficient ($d_{33}^{effective}$) of ~70 pm/V was extracted for devices with 70 μm electrode diameters for PZT on templated Si and ~54 pm/V for PZT on Pt-Si. The larger response for the BTO templated stack is consistent with the orientation dependence of PZT thin films, in which well-textured (001) oriented films exhibit higher longitudinal response than mixed or (111) textured films [19]. Comparable orientation-controlled enhancements of piezoelectric response on oxide-buffered Si have been reported for epitaxial PZT devices [40]. To assess electromechanical durability, the devices were subjected to cyclic bipolar electric fields with $V_{max}$ of 10 (V. The corresponding strain response showed no signs of degradation up to $10^6$ cycles, indicating excellent fatigue resistance and suitability for micro and nanoelectromechanical system (MEMS/NEMS) applications.

## 3 Conclusions

In summary, we report a hybrid integration route for high-quality ferroelectric oxide films on Si, enabled by the transfer of thin epitaxial BTO membranes and subsequent chemical-solution deposition (CSD) of PZT. A KI + HCl redox etch significantly accelerates dissolution of the SVO sacrificial layer, enabling full membrane release in about 30 minutes, reducing the release time from days to minutes compared with conventional water-based release of $AVO_3$/$AMoO_3$ (A = Ca, Sr, Ba). The transferred BTO template facilitates cube-on-cube epitaxy during the crystallization of CSD-grown PZT, yielding films with robust ferroelectric performance.

Electrical measurements show a remanent polarization of 10–12 µC/cm² and a coercive field of ~100 kV/cm. Endurance testing confirms stable switching to $10^8$ cycles on BTO VS, whereas control devices on Pt–Si exhibit fatigue beyond ~$10^6$ cycles. Laser Doppler vibrometry reveals a stronger longitudinal electromechanical response for devices on BTO VS and no measurable degradation of the butterfly loop over $10^6$ cycles. These results underscore the viability of transferred oxide templates for integrating functional ferroelectrics on Si while avoiding high-temperature oxide epitaxy directly on Si.

Importantly, the CSD route offers a cost-effective, scalable alternative to vacuum deposition with minimal equipment complexity. The overall transfer and deposition flow is chemically versatile and scalable in concept. Future efforts will focus on interfacial engineering, process window optimization, and scaling to larger substrates, with an eye toward pre-metal (front-end) integration of ferroelectric and piezoelectric oxides in CMOS-adjacent platforms.

**4 Experimental Section**

*Epitaxial thin film growth:* Thin films were grown using pulsed laser deposition (PLD) utilizing an excimer laser with 248 nm wavelength. Single-crystal $SrTiO_3$ (001) substrates were cleaned by sequential sonication in acetone, isopropanol, and deionized water, then dried under $N_2$. $SrVO_3$ (SVO) sacrificial layers films were deposited by pulsed-laser deposition on $SrTiO_3$ (001) with laser fluence of 1.4 J/cm², a base pressure of $3 \times 10^{-6}$ mbar, and a substrate temperature of 750 °C. $BaTiO_3$ (BTO) films were deposited by pulsed-laser deposition on SVO/STO with laser fluence of 1.2 J/cm², a base pressure of $5 \times 10^{-6}$ mbar, and a substrate temperature of 800 °C. Reflection high energy electron diffraction (RHEED) was employed to monitor SVO and BTO growth on STO. SAO films were also grown on STO under a laser fluence of 1.3 J/cm², at 800 °C and $5 \times 10^{-6}$ mbar, followed by BTO deposition under the same conditions used above. The BTO/SAO//STO heterostructure was subsequently annealed at 900 °C.

*Transfer and Characterization of Freestanding Membranes:* A polymer support (PMMA) was spin-coated on BTO/SVO//STO and soft-baked (120 °C, 2 min). Samples were immersed at room temperature in an aqueous KI + HCl etchant to dissolve SVO. Complete lift-off occurred in about 30 minutes, monitored optically. Released BTO membranes were rinsed in deionized water and isopropanol, dried, and scooped onto Pt-coated Si wafers (Pt/Ti/$SiO_2$/Si). The polymer support was removed in solvent, followed by a mild bake (150–200 °C) to promote adhesion. AFM images were taken using a Park AFM system operating in tapping mode, and X-ray diffraction data were obtained using a monochromate Cu-K$\alpha$1 source on a Rigaku four-circle machine.

*Chemical-Solution Deposition (CSD) of PZT:* A 0.52/0.48 PZT precursor solution was filtered (0.2 µm) and spin-coated (4000 rpm, 30 s) onto BTO/Pt–Si and, for controls, directly onto Pt–Si. Each layer was solvent-flashed (150 °C, 1 min), pyrolyzed (350–400 °C, 2 min), and crystallized at 650 °C (rapid thermal step ~2 min followed by a 15 min hold) in oxygen-containing ambient. This cycle was repeated to reach the target thickness (~250 nm)

*Device Fabrication:* Circular tungsten (W) top electrodes were defined by photolithography and lift-off after DC sputter deposition of W (~100 nm). Electrode diameters used for electrical measurements were 70 µm. The bottom electrode was the continuous Pt layer on Si.

*STEM Imaging and Analysis*: Cross-sectional lamellae were prepared by focused ion-beam milling (FEI Scios 2). STEM imaging was performed on an aberration-corrected Titan Themis operated at 300 kV (probe convergence 24 mrad; HAADF collection 48–200 mrad). Compositional mapping used a non-aberration-corrected Titan Themis (300 kV) equipped with a ChemiSTEM EDS system; STEM-EDS data were integrated from binned pixels until a signal-to-noise ratio > 5 was reached.

*Ferroelectric and electromechanical response:* Parallel-plate capacitors with top–bottom electrodes were characterized at room temperature. Polarization vs voltage loops, instantaneous current vs voltage, and fatigue measurements were obtained using a Radiant Technologies Precision Multiferroic tester. Out-of-plane displacement was measured by laser Doppler vibrometry (MSA 500) in double-beam interferometry mode with 532 nm reference and probe beams. Displacement waveforms were recorded over 30–40 ms and ensemble-averaged to improve signal-to-noise.

**Associated Contents**

**Supporting Information**

**Author Information**

Corresponding Authors

*Pavan Nukala, Email : pnukala@iisc.ac.in

*Asraful Haque, Email : asrafulhaque@iisc.ac.in

Present Addresses

*Center for Nanoscience and Engineering, Indian Institute of Science, Bengaluru, India, 560012

Notes

The authors declare no competing financial interest.

**Acknowledgements**
This work was partly carried out at the Micro and Nano Characterization Facility (MNCF) and National Nanofabrication Center (NNfC) located at CeNSE, IISc Bengaluru and benefitted from all the help and support from the staff. The authors further acknowledge the advanced

## Supporting Information (SI)

## A facile vector substrate platform via BaTiO₃ membrane transfer enables high quality solution processed epitaxial PZT on silicon

Asraful Haque*, Antony Jeyaseelan, Shubham Kumar Parate, Srinivasan Raghavan, Pavan Nukala*.
Centre for Nanoscience and Engineering, Indian Institute of Science, Bengaluru, India, 560012.

### Redox assisted dissolution of SrVO₃ in KI and HCl

In acidic chloride media (pH ≤ 2), the iodide–iodine can mediate oxidative dissolution at the perovskite surface, while HCl provides both acidity and chloride for complexation.

$$I_2 + 2e^- \leftrightarrows 2I^- \quad (E^o \approx +0.535 - 0.536 \, V \, vs \, SHE \, at \, 25 \, °C) \; [1]$$

Under these conditions, iodide/iodine act as a redox shuttle and chloride stabilizes soluble vanadium species; the dominant aqueous V species in strongly acidic solution includes: $VO^{2+}$ ($V^{4+}$, vanadyl), $VO_2^+$ ($V^{5+}$, di-oxovanadium), and chloro-(oxo)-vanadates (e.g., $VO_2Cl(aq)$, $VOCl_4^-$, $VOCl_5^{2-}$), with distributions depending on pH, redox potential, and chloride activity [2].

An illustrative (non-unique) net oxidative dissolution step is:

$$2SrVO_3(S) + 4H^+ + I_2 \rightarrow 2Sr^{2+} + 2VO^{2+} + 2H_2O + 2I^-$$

This equation is schematic and provided for guidance; the actual distribution among $VO^{2+}$, $VO_2^+$, and chloro-(oxo)-vanadate complexes will vary with solution composition (pH, [Cl⁻], and oxidant activity) [2].

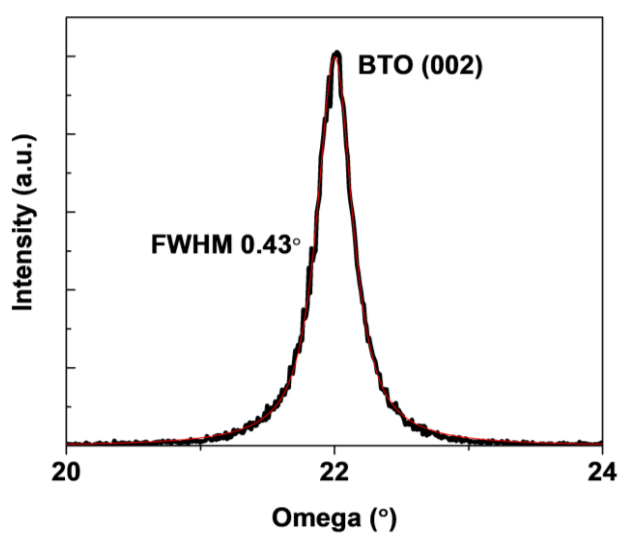

**Figure S1:** Rocking curve around the BTO (002) reflection of as grown BTO on SVO//STO.

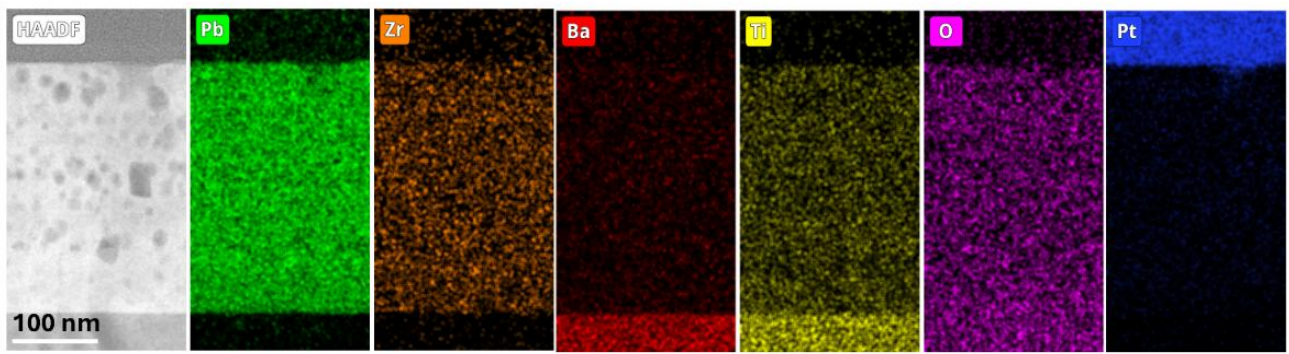

**Figure S2:** STEM-EDS shows the elemental mapping of Pb, Zr, Ba, Ti, O, and Pt in the PZT, BTO and top Pt regions in Pt/PZT/BTO//Pt-Si.

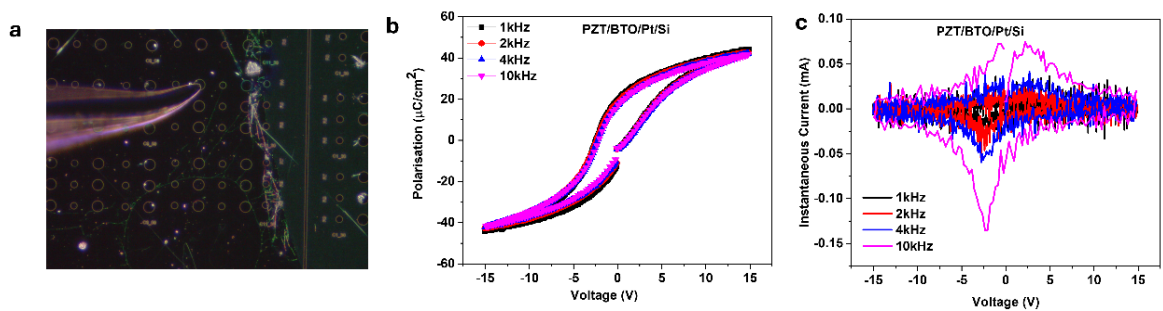

**Figure S3:** a) Optical image of circular W electrodes fabricated on the PZT grown on BTO VS (b) Polarization–voltage hysteresis loop measured at different frequencies. (c) Instantaneous current (I) vs voltage (V) response corresponding to polarization switching.